\documentclass[12pt]{article}

\newcommand{\beq}{\begin{equation}}
\newcommand{\beql}[1]{\begin{equation}\label{eq:#1}}
\newcommand{\eeq}{\end{equation}}
\newcommand{\be}{\begin{equation}}
\newcommand{\ee}{\end{equation}}
\newcommand{\beqn}{\begin{eqnarray}}
\newcommand{\eeqn}{\end{eqnarray}}
\newcommand{\bea}{\begin{eqnarray}}
\newcommand{\eea}{\end{eqnarray}}

\newcommand{\eq}[1]{(\ref{eq:#1})}

\DeclareFixedFont{\xiiss}{OT1}{cmss}{m}{n}{12}
\DeclareFixedFont{\ixss}{OT1}{cmss}{m}{n}{9}
\DeclareFixedFont{\cmrnine}{OT1}{cmr}{m}{n}{9}

\newcommand{\CC}{\hbox{\xiiss C\kern-.4emI}}
\newcommand{\RR}{\hbox{\xiiss R\kern-.45emI}}
\newcommand{\ZZ}{\hbox{\xiiss Z\kern-.4emZ}}
\newcommand{\CCs}{\hbox{\ixss C\kern-.4emI}}
\newcommand{\ZZs}{\hbox{\ixss Z\kern-.4emZ}}
\newcommand{\pa}{\partial}
\newcommand{\half}{{\scriptscriptstyle {1\over2}}}
\newcommand{\tr}{{\rm tr}\ }
\newcommand{\pasl}{\pa\kern-.55em /}
\newcommand{\Dsl}{D\kern-.65em /}

\begin{document}
\begin{titlepage}
\title{
	\begin{flushright}
	\begin{small}
	ILL-(TH)-98-01\\
	hep-th/9803068\\
	\end{small}
	\end{flushright}
	\vspace{1.cm}
The  Large $N$ Limit of the $(2,0)$ Superconformal  Field Theory
}

\author{
Robert G. Leigh\thanks{U.S. Department of Energy Outstanding Junior Investigator}
\thanks{e-mail: \tt rgleigh@uiuc.edu}
\ and
Moshe Rozali\thanks{e-mail: \tt rozali@hepux1.hep.uiuc.edu}\\
\\
	\small\it Department of Physics\\
	\small\it University of Illinois at Urbana-Champaign\\
	\small\it Urbana, IL 61801
}

\maketitle

\begin{abstract}
 
We discuss the large $N$ limit of the $(2,0)$ field theory in six
dimensions. We do this by assuming the validity of Maldacena's
conjecture of the correspondence between  large $N$ gauge theories and
supergravity backgrounds, here   $AdS_7\times S^4$. We review the spectrum of
the supergravity theory and compute the  spectrum of primary
operators of the conformal algebra of arbitrary spin.

\end{abstract}

\end{titlepage}

%%%%%%%%%%%%%%%%%%%%%%%%%%%%%%%%%%%%%%%%%%%%%%%%%%%%%%%%%%%%%%%%%%%%%%%%%%%%%

\section{Introduction}

An interesting development stemming from investigations of string
duality is the existence of higher dimensional interacting
superconformal fixed points. The simplest such fixed point is the $(2,0)$
field theory in six dimensions. The theory can be discovered on the
world-volume of the M-theory 5-branes \cite{stro}, and in the
compactification of type IIB string theory on a singular $K3$
\cite{witten}.  This theory, compactified on a 5-torus, is also relevant
to Matrix theory \cite{bfss} compactifications on $T^4$ \cite{roz, brs}.

The realization of the theory as embedded in  M-theory leaves many questions open.
The theory of a single fivebrane is a free field theory containing one
free tensor multiplet.  For $N$ fivebranes one expects to get an
interacting field theory at a fixed point of the renormalization group
\cite{sei16}. One would like therefore to compute dimensions of primary
operators and their correlation functions in order to extract
information about the theory.

Some progress has been made in Ref. \cite{abs}, in which the DLCQ of the
$(2,0)$ theory was discussed, via the Matrix description of M-theory.
Several quantities were computed, for low values of $N$.

Recently, Maldacena \cite{mald3,maldn} has made the suggestion that large
$N$ gauge theories are related to supergravity on 
anti-deSitter ($AdS$) space.\footnote{For examples of related work, see \cite{premald}.} In
the case of maximal supersymmetry, the full space is $AdS_n\times C$,
where $C$ is a sphere of appropriate dimension. This conjecture has been
significantly clarified by the recent work of \cite{witads}. In
the present paper, we study the case of supergravity on $AdS_7\times
S^4$, which is related to the theory of the M5-brane, and thus should
give information on the $(2,0)$ theory in the large $N$ limit. Similar
studies have been done \cite{ferraras,krn} in the $AdS_5\times S^5$ case,
where the corresponding gauge theory is the relatively well-understood
$N=4, d=4$ $SU(N)$ Yang-Mills theory. In this case, however, we expect to
learn something new about the $(2,0)$ theory, at least in the infinite 
$N$ limit.

Assuming the
validity of the conjecture, there is a correspondence between the
spectrum of masses on the AdS space and the spectrum of dimensions of
operators in the conformal field theory \cite{witads}. In section 2 we
extend the analysis of \cite{witads} to non-scalar operators. In
particular we find a correlation  between the sign of the mass of a
fermionic state on $AdS_7$ and the chirality of the operator it induces on
the boundary.
Using this analysis we construct in section 3 the spectrum of the
conformal primary operators of arbitrary spin in the (2,0) theory.
According to the conjecture in \cite{mald3} this is a complete list of
the primary operators whose dimension is finite in the infinite $N$
limit. 

After this work was completed, we received two papers which discuss
similar issues \cite{aoy,minnew}. Our results, where they overlap, are in
agreement.

\section{KK Modes of $d=11$ Supergravity on $S^4$}

In \cite{witads}  an interpretation of Maldacena's conjecture was
suggested. The gauge theory is described by the supergravity background
via a holographic image on the boundary of the $AdS$ space. There is a
one-to-one mapping of bulk supergravity modes and conformal operators in
the boundary theory. The Kaluza-Klein modes on the sphere play a special
role, since they describe the complete set of operators whose dimensions
are independent of $N$ in the large $N$ limit. It is known that all such
Kaluza-Klein modes appear as short multiplets of the supersymmetry
algebra in an $AdS$ background \cite{stab}.

The scaling dimension of a given boundary operator is  related to a mass
parameter of the corresponding bulk solution. In this section, we review
this relation and extend it to operators of arbitrary spin. The Kaluza-Klein
modes of $d=11$ supergravity on $S^4$ are known, having been worked out
in Refs. \cite{gnw,nieu}. It should thus be a simple task to identify
the primary operators of the $d=6$ superconformal theory. The operators, 
being elements of short multiplets, should saturate certain
unitarity bounds which have recently been worked out in Ref.
\cite{minwalla}.

We will refer to the bulk as $M$, and the 6-dimensional boundary as $B$.
We denote a mode in the bulk generically as $j\{\alpha\}$, where
$\alpha$ is a collection of quantum numbers labeling the representation
of the superconformal algebra $Sp(6,2|4)$; it is sufficient to give
quantum numbers of the maximal compact subgroup, here $Spin(5,1)\times
Spin(2)\times Sp(2)_R$. The bulk mode couples to an operator on the
boundary ${\cal O}\{\bar\alpha\}$, where $\bar\alpha$ is the conjugate
representation. For example, if $j$ is a
chiral spinor, the corresponding ${\cal O}$ is an anti-chiral spinor;
similarly if $j$ is a self-dual tensor, then ${\cal O}$ is
anti-self-dual. The coupling is simply of the form
\beq 
\int_B j\{\alpha\} \cdot {\cal O}\{\bar\alpha\} 
\eeq

\subsection{Dimension formulas}

The component fields of the supergravity solutions on $AdS_7\times S_4$
appear in Table 1 of Ref. \cite{gnw}. All the rows in that table
are determined from the first by supersymmetry and group theory; however,
it is interesting to see how this works in detail 
from the point of view of supergravity solutions for the various 
component fields. In what follows we analyze the behaviour
of the 7-dimensional solutions at the boundary. Throughout, we
work in the metric
\beq
ds^2=dy^2+\sinh^2 y\ \eta_{ij} dx^i dx^j
\eeq
The boundary is at $y\to\infty$.

\bigskip
\underline{Scalars}: 
As given in \cite{witads}, scalar solutions behave as $e^{\lambda y}$ at
infinity, and the Klein-Gordon equation for the scalar in this background
implies $\lambda(\lambda+d)=m^2$. The dominant root is the larger
of the two, and is generically positive.\footnote{For the rest of the
Lorentz representations, we also take the corresponding $\lambda$ to be
positive.} The corresponding operator in the SCFT then has dimension
$\Delta=\lambda+d$. There are three series:\footnote{In Table 1 of 
Ref. \cite{gnw}, masses should be scaled by a factor of $e=2$.}
\beqn
m^2=4k(k-3):&&\ \ \Delta=2k, \ \ (k=1,2,\ldots )\\
m^2=4(k^2+7k+10):&&\ \ \Delta=2k+10, \ \ (k=0,1,\ldots)\label{eq:scaltwo}\\
m^2=4(k^2+9k+18):&&\ \ \Delta=2k+12, \ \ (k=0,1,\ldots)\label{eq:scalthr}
\eeqn
As we will see below, in a generic massive multiplet, there are scalar 
components which fit into each of these series. The first few multiplets
have scalars only from the first series.

\bigskip
\underline{Spinors}: 
Here, we should study solutions of the Dirac equation in the
AdS background at infinity to find the dimensions of the corresponding
operators. After a brief calculation, we find that there is an important 
contribution from the spin connection in this background:
\beq
\left(\Dsl_7+m\right)\Psi=\left[\gamma^7\left(\pa_y+d/2 \coth y\right)
+m+{1\over\sinh y}\pasl_6\right]\Psi
\eeq
and so at infinity
\beq
\Psi\sim e^{\lambda y} u_{\infty}\\
\eeq
\beq
\left[ (\lambda+d/2)\gamma^7+m\right] u_\infty=0
\eeq
In the $5+1$-dimensional sense, $\gamma^7$ gives the chirality. Since
$\lambda$ is taken to be positive, this correlates the sign of the mass
with chirality. Hence:
\beqn
m>0:\ \ \Delta_-=+m+d/2\\
m<0:\ \ \Delta_+=-m+d/2
\eeqn
This is an important result, and as we shall see in the next section,
precisely what is needed for the interpretation in terms of operators
in the $(2,0)$ theory.

From table 1 of \cite{gnw} (shifting range of $k$ appropriately in some cases), 
there are four series of spinors:
\beqn
m_1=e(k-1/4)\ \ \ && \Delta_-=2k+5/2\ \ (k=0,1,\ldots )\\
m_2=-e(k+1/4)\ \ \ && \Delta_+=2k+15/2\ \ (k=0,1,\ldots )\\
m_3=e(k+3/4)\ \ \ && \Delta_-=2k+21/2\ \ (k=0,1,\ldots)\\
m_4=-e(k+5/4)\ \ \ && \Delta_+=2k+23/2\ \ (k=0,1,\ldots )
\eeqn
with $5+1$-dimensional chirality indicated as before.

One should note that in the corresponding 4-dimensional $N=4$ case, obtained
from Type IIB on $AdS_5\times S_5$, we
should interpret the spinors with definite scaling as $d=5$
symplectic-Majorana spinors and the mass as a Majorana mass. In this
case, for each mode in the bulk, we find an equation like
$\gamma^7(\lambda_i+d/2)u_{i,\infty}= m\epsilon_{ij}u_{j,\infty}$. 
 Choosing  $\lambda_i$ positive, we get two Weyl spinors of
opposite chirality, with the same dimension. The spectrum is then 
non-chiral, as it must be to have $N=4$ supersymmetry on the boundary.

\bigskip
\underline{Vectors and $p$-forms}: Here we are solving an equation of the form
\beql{vecwave}
\pa_\mu\left( \sqrt{-g}g^{\mu\nu}g^{\rho\sigma}F_{\nu\sigma}\right)
=m^2\sqrt{-g}g^{\rho\mu}A_\mu
\eeq
To see in detail the behaviour at infinity, we choose a convenient gauge
$\nabla_{(6)}\cdot A=0$, $A_7=0$. The remaining components of eq. \eq{vecwave}
are now
\beql{vecredw}
\pa_\ell \left(\sqrt{-g}\ g^{\ell j}g^{ki} F_{ji}\right)+
\pa_y \left(\sqrt{-g}\ g^{77}g^{ki} F_{7i}\right)
=m^2\sqrt{-g}\ g^{ki} A_i
\eeq
At infinity, if $A_i\sim e^{\lambda y}f_i(x)$, the first factor in \eq{vecredw}
is subleading, and we find
\beq
\lambda (\lambda+d-2) = m^2
\eeq
It is important to realize that there is a factor of $g^{77}$
in the second factor of \eq{vecredw} in doing the power counting.

For $p$-form gauge fields satisfying a similar Maxwell-like equation, the
result generalizes to
\beq
\lambda(\lambda+d-2p)=m^2
\eeq
The dimension of the corresponding operator has \cite{witads} 
$\Delta=-p+\lambda+d$, and so we get
\beq
(\Delta+p-d)(\Delta-p)=m^2
\eeq
As an example, we note that for the first vector series, we have $m^2=4(k^2-1)$
and so $(\Delta-5)(\Delta-1)=4k^2-4$, which gives $\Delta_{v1}=3+2k, k=1,2,3,\ldots$. For the
first 3-form series, we have $m^2=4k^2$, and so $(\Delta-3)^2=4k^2$, i.e. $\Delta_{T1}
=3+2k, k=0,1,2,\ldots$. All other $p$-form series may be found in the Table
at the end of this paper.

\bigskip
\underline{Graviton}: For spin 2 fields, the equations of motion can be
shown to reduce to the scalar case, and thus $\Delta(\Delta-d)=m^2$. The
correct value of $m^2$ is $4(k^2+3k)$ (a shift of $-1/2$ appeared incorrectly
(for the notation used here) in Table 1 of eq. \cite{gnw}.) Thus $\Delta=6+2k,
\ k=0,1,2,\ldots$.

\bigskip
\underline{Gravitini}: Here, we repeat a similar analysis as for the spinors.
The result is the same and again the results are chiral.

\section{ The Spectrum of Operators}

An operator in the (2,0) theory is characterized by Lorentz quantum
numbers and by the $Sp(2)_R$ R-symmetry representation. Let the highest
weights of the operators under $Spin(5,1)$ be $(h_1,h_2, h_3)$ with
index $T(h)$, and the highest weights under $Sp(2)_R$ be
$(\ell_1, \ell_2)$.

A representation of the superconformal algebra may be constructed by
starting with a superconformal primary operator, which by definition is
annihilated by the special conformal generators $K_i$ and by the special
SUSY generators $S_{\alpha}$. Any superconformal descendant is
constructed by acting with ``raising'' operators which are the SUSY
generators $Q_{\alpha}$ and the momentum operators $P_i$.

A superconformal primary operator is called of level $r$ if acting with
the raising operators $r$ times we encounter for the first time a null
state. Such primary operators generate short representations of the
superconformal algebra, and saturate unitarity bounds derived in
\cite{minwalla}.  The dimension, $\Delta$, of such an operator is
specified by its quantum numbers as follows:

For  level 1 operators one has:
\begin{equation}
\Delta = T(h) + 2 (\ell_1+\ell_2)
\end{equation}

For level 2 operators there are 3 possibilities:
\begin{equation}
\Delta = T(h) + 2 (\ell_1 + \ell_2 ) + a
\end{equation}  
where $a$ can be 2,4 or 6.

Given that the Kaluza Klein modes are in short multiplets, we expect
them to saturate one of the bounds above. Indeed all operators are in
fact level one.

It turns out that we can classify, at least in terms of $Spin(6)\times
Sp(2)_R$ quantum numbers, the gauge invariant primary operators that
appear in the superconformal theory in terms of place-holder fields
\beq
\phi,\ \psi,\ H
\eeq
all of which are taken to be in the adjoint representation of $U(N)$ and
transform as a tensor multiplet of the $(2,0)$ supersymmetry. The
$Spin(6)$ weights are respectively
\beq
(0,0,0),\ (\half,\half,\half),\ (1,1,1).
\eeq
Similarly, the $Sp(2)_R$ weights are
\beq
(1,0),\ (\half,\half),\ (0,0).
\eeq

The primary operators may then simply be identified with
\beql{place}
{\cal O}_{m,n,k}=\tr H^m \psi^n \phi^k
\eeq
A hierarchy is built up based on the integer $p=m+n+k=0,1,2,\ldots$.
Clearly the $p=0$ operator is just the identity operator, with dimension
zero. It is a singlet of the supersymmetry algebra. The $p=1$ operators
$\tr H$, $\tr\psi$, $\tr\phi$ are in $Sp(2)_R$ representations ${\bf
1}$, ${\bf 4}$, ${\bf 5}$ respectively. Clearly, they touch only the
Abelian part of $U(N)$; they correspond to the doubleton and have 
free dimensions $3,5/2,2$ respectively.

In the bulk supergravity theory, the doubleton representation corresponds
to decoupled, pure gauge modes. In the boundary theory, they represent a
free Abelian tensor multiplet which decouples from the interacting fixed
point theory. Note that we build all representations from the place-holder fields \eq{place}.
These should not be confused with the doubleton representation.

We collect the results of Section 2 in a Table below. All of the Kaluza-Klein
modes give rise to operators \eq{place} whose dimensions can be read off by
simply adding the dimensions of the place-holder fields. All the superconformal
primary fields saturate the level 1 unitarity bounds. We note also for example
that the additional series of scalars \eq{scaltwo} and \eq{scalthr} appear
as scalar "composites" of the place-holder fields. Furthermore, the chirality
of the various spinor series is determined by the $Spin(6)$ group theory, and
agrees with the supergravity analysis of the last section.

The structure of the table, and in particular the ability to represent
all operators as simple composites of a set of place-holder fields, is
similar to the structure of a free field theory. Indeed, far along the
flat directions of the (2,0) field theory, the infra-red physics is
free. The eigenvalues of the place-holder fields are then interpreted as
the $N$ free tensor multiplets, whose scalars parametrize the moduli
space $\RR^{5N}/S_N$. As we move towards the singularities of the moduli
space, we expect to reach an interacting (2,0) superconformal fixed
point. A general operator will undergo significant renormalization,
unless it is protected by SUSY non-renormalization theorems.

For this reason, not all possible combinations of the fields $\phi,
\psi, H $ appear in the table.  The only superconformal primary fields
appearing in the table correspond to the completely symmetric
combination of the $\phi$ fields.  For the completely symmetric
combination, the R-symmetry weights $\ell_1, \ell_2$ simply add up.
Therefore the unitarity bounds given above completely determine the
dimension of these operators (and therefore also their descendants). In
other words the operators found in the table are precisely those which
are protected from renormalization.

Other combinations of $\phi$'s, involving their commutators, are not
constrained by this argument to saturate the unitarity bound. Such
operators, since not protected by non-renormalization theorems, are
expected to have dimensions which diverge with $N$, and indeed we find
that they do. In the context of Maldacena's conjecture, they correspond
to M-theory modes which are not seen in the supergravity approximation.

Work supported in part by DOE grant DE-FG02-91ER40677. 

\pagebreak

\bigskip
\begin{center}
\begin{tabular}{|c|c|l|l|}					\hline
Spin(6)		&Sp(2)$_R$	&	dimension		&	Operator	\\	\hline\hline
{\bf 1}		&{\bf 5}	&	0 + 2(1)		&$\tr\phi$		\\	\hline
{\bf 4}		&{\bf 4}	&	$5/2$ + 2(0)	&$\tr\psi$		\\	\hline
{\bf 10}	&{\bf 1}	&	3 + 2(0)		&$\tr H$		\\	\hline\hline
{\bf 1}		&{\bf 14}	&	0 + 2(2)		&$\tr\phi\phi$
	\\	\hline
{\bf 4}		&{\bf 16}	&	$5/2$ + 2(1)	&$\tr\psi\phi$	\\	\hline
{\bf 10}	&{\bf 5}	&	3 + 2(1)		&$\tr H\phi$	\\	\hline
{\bf 6}	&{\bf 10}	&	5 + 2(0)			&$\tr\psi\psi$ \\	\hline
{\bf 20}	&{\bf 4}	&	11/2 + 2(0)		&$\tr\psi H$	\\	\hline
{\bf 20'}	&{\bf 1}	&	6 + 2(0)		&$\tr HH$
		\\	\hline\hline
{\bf 1}		&{\bf 30}	&	0 + 2(3)		&$\tr\phi\phi\phi$		\\	\hline
{\bf 4}		&{\bf 40}	&	$5/2$ + 2(2)	&$\tr\psi\phi\phi$		\\	\hline
{\bf 10}	&{\bf 14}	&	3 + 2(2)		&$\tr H\phi\phi$		\\	\hline
{\bf 6}	&{\bf 35}	&	6 + 2(1)			&$\tr\psi\psi\phi$		\\	\hline
{\bf 20}	&{\bf 16}	&	11/2 + 2(1)		&$\tr\psi H\phi$	\\	\hline
{\bf 20'}	&{\bf 5}	&	6 + 2(1)		&$\tr HH\phi$\\	\hline
${\bf \overline{ 4}}$	&{\bf 20}	&	$15/2$ + 2(0)	&$\tr\psi\psi\psi$		\\	\hline
${\bf 15}$	&{\bf 10}	&	8 + 2(0)		&$\tr\psi\psi H$		\\	\hline
${\bf \overline{ 20}}$	&{\bf 4}	&	$17/2$ + 2(0)	&$\tr\psi HH$		\\	\hline
{\bf 10}&{\bf 1}	&	9 + 2(0)			&$\tr HHH$				\\	\hline\hline
{\bf 1}		&{\bf 55}	&	0 + 2(4)		&$\tr\phi\phi\phi\phi$		\\	\hline
{\bf 4}		&{\bf 80}	&	$5/2$ + 2(3)	&$\tr\psi\phi\phi\phi$		\\	\hline
{\bf 10}	&{\bf 30}	&	3 + 2(3)		&$\tr H\phi\phi\phi$		\\	\hline
{\bf 6}	&{\bf 81}	&	6 + 2(2)			&$\tr\psi\psi\phi\phi$		\\	\hline
{\bf 20}	&{\bf 40}	&	11/2 + 2(2)		&$\tr\psi H\phi\phi$	\\	\hline
{\bf 20'}	&{\bf 14}	&	6 + 2(2)		&$\tr HH\phi\phi$\\	\hline
${\bf \overline{ 4}}$	&{\bf 64}	&	$15/2$ + 2(1)	&$\tr\psi\psi\psi\phi$		\\	\hline
${\bf 15}$	&{\bf 35}	&	$8$ + 2(1)		&$\tr\psi\psi H\phi$		\\	\hline
${\bf \overline{ 20}}$	&{\bf 16}	&	$17/2$ + 2(1)	&$\tr\psi HH\phi$		\\	\hline
{\bf 10}&{\bf 5}	&	9 + 2(1)			&$\tr HHH\phi$				\\	\hline
{\bf 1}		&{\bf 35}	&	10 + 2(0)		&$\tr\psi\psi\psi\psi$		\\	\hline
${\bf \overline{4}}$	&{\bf 20}	&	$21/2$ + 2(0)	&$\tr\psi\psi\psi H$		\\	\hline
{\bf 6}	&{\bf 10}	&	11 + 2(0)			&$\tr\psi\psi HH$ \\	\hline
{\bf 4}		&{\bf 4}	&	$23/2$ + 2(0)	&$\tr\psi HHH$		\\	\hline
{\bf 1}&{\bf 1}	&	12 + 2(0)				&$\tr HHHH$				\\	\hline\hline
\end{tabular}
\end{center}
{{\bf Table:} The first four superconformal multiplets (multiplets are separated
by a double line). All other multiplets are obtained by appending extra $\phi$'s
in the traces of the last multiplet. Throughout the table, dimensions are given
in the form $n+2(k)$, where $k$ counts the number of $\phi$'s in the given
operator. The  R-symmetry index structure of the operators  is supressed and can
be recovered from the corresponding representation.}

%\bibliography{matrix6}
\bibliographystyle{unsrt}

\def\npb#1#2#3{Nucl. Phys. {\bf B#1} (#2) #3}
\def\plb#1#2#3{Phys. Lett. {\bf #1B} (#2) #3}
\def\prd#1#2#3{Phys. Rev. {\bf D#1} (#2) #3}
\def\prl#1#2#3{Phys. Rev. Lett. {\bf #1} (#2) #3}
\def\cqg#1#2#3{Class. Quantum Grav. {\bf #1} (#2) #3}
\def\mpl#1#2#3{Mod. Phys. Lett. {\bf A#1} (#2) #3}
\def\hepth#1#2#3#4#5#6#7{hep-th/#1#2#3#4#5#6#7}

\end{document}